\documentclass[sigconf]{acmart}

\setcopyright{none} %
\acmConference[Accepted Submission to ACM CCS 2017 (PREPRINT)]{ACM Conference on Computer and Communications Security}{Due 19 May 2017}{Dallas, Texas}
\acmYear{2017}

\usepackage{times}
\usepackage{pifont} 
\usepackage{subfigure}
\usepackage{url}
\usepackage{breakurl}
\usepackage{paralist}
\usepackage{booktabs}
\usepackage{listings}
\usepackage{graphicx}
\usepackage{enumitem}
\usepackage{pbox}
\usepackage{arydshln}
\usepackage{amssymb}
\usepackage[font={small}]{caption}

\newcommand{\cyborg}{HaCRS}
\newcommand{\binarycount}{85}

\begin{document}

\title{Rise of the HaCRS:\\\large{Augmenting Autonomous Cyber Reasoning Systems with Human Assistance}}

\author{Yan Shoshitaishvili}
\affiliation{\institution{Arizona State University}}
\email{yan.shoshitaishvili@asu.org}

\author{Michael Weissbacher}
\affiliation{\institution{Northeastern University}}
\email{mw@ccs.neu.edu}

\author{Lukas Dresel}
\affiliation{\institution{UC Santa Barbara}}
\email{lukas.dresel@cs.ucsb.edu}

\author{Christopher Salls}
\affiliation{\institution{UC Santa Barbara}}
\email{salls@cs.ucsb.edu}

\author{Ruoyu Wang}
\affiliation{\institution{UC Santa Barbara}}
\email{fish@cs.ucsb.edu}

\author{Christopher Kruegel}
\affiliation{\institution{UC Santa Barbara}}
\email{chris@cs.ucsb.edu}

\author{Giovanni Vigna}
\affiliation{\institution{UC Santa Barbara}}
\email{vigna@cs.ucsb.edu}

\date{}

\begin{abstract}
Software permeates every aspect of our world, from our homes to the infrastructure that provides mission-critical services.

As the size and complexity of software systems increase, the number and sophistication of software security flaws increase as well.
The analysis of these flaws began as a manual approach, but it soon became apparent that a manual approach alone cannot scale, and that tools were necessary to assist human experts in this task, resulting in a number of techniques and approaches that automated certain aspects of the vulnerability analysis process. 

Recently, DARPA carried out the Cyber Grand Challenge, a competition among autonomous vulnerability analysis systems designed to push the tool-assisted human-centered paradigm into the territory of complete automation, with the hope that, by removing the human factor, the analysis would be able to scale to new heights.
However, when the autonomous systems were pitted against human experts it became clear that certain tasks, albeit simple, could not be carried out by an autonomous system, as they require an understanding of the logic of the application under analysis. 

Based on this observation, we propose a shift in the vulnerability analysis paradigm, from tool-assisted human-centered to human-assisted tool-centered. 
In this paradigm, the automated system orchestrates the vulnerability analysis process, and leverages humans (with different levels of expertise) to perform well-defined sub-tasks, whose results are integrated in the analysis.
As a result, it is possible to scale the analysis to a larger number of programs, and, at the same time, optimize the use of expensive human resources.

In this paper, we detail our design for a human-assisted automated vulnerability analysis system, describe its implementation atop an open-sourced autonomous vulnerability analysis system that participated in the Cyber Grand Challenge, and evaluate and discuss the significant improvements that non-expert human assistance can offer to automated analysis approaches.

\end{abstract}

\maketitle

\section{Introduction}

Software has become dominant and abundant.
Software systems support almost every aspect of our lives, from health care to finance, from power distribution to entertainment.
This growth has led to an explosion of software bugs and, more importantly, software vulnerabilities.
Because the exploitation of vulnerabilities can have catastrophic effects, a substantial amount of effort has been devoted to discovering these vulnerabilities before they are found by attackers and exploited in the wild.

Traditionally, vulnerability discovery has been a heavily manual task.
Expert security researchers spend significant time analyzing software, understanding how it works, and painstakingly sifting it for bugs.
Even though human analysts take advantage of tools to automate some of the tasks involved in the analysis process, the amount of software to be analyzed grows at an overwhelming pace.
As this growth reached the scalability limits of manual analysis, the research community has turned its attention to \emph{automated program analysis}, with the goal of identifying and fixing software issues on a large scale.
This push has been met with significant success, culminating thus far in the DARPA Cyber Grand Challenge (CGC)~\cite{darpacgc}, a cyber-security competition in which seven finalist teams pitted completely autonomous systems, utilizing automated program analysis techniques, against each other for almost four million dollars in prize money.

By removing the human factor from the analysis process, the competition forced the participants to codify the strategy and orchestration tasks that are usually performed by experts, and, at the same time, it pushed the limits of current vulnerability analysis techniques to handle larger, more complex problems in an efficient and resource-aware manner. 
These systems represented a significant step in automated program analysis, automatically identifying vulnerabilities and developing exploits for 20 of a total of 82 binary programs developed for the event.

Despite the success of these systems, the underlying approaches suffer from a number of limitations.
These limitations became evident when some of the CGC autonomous systems participated in a follow-up vulnerability analysis competition (the DEFCON CTF) that included human teams.
The autonomous systems could not easily understand the logic underlying certain applications, and, as a result, they could not easily produce inputs that drive them to specific (insecure) states.
However, when humans could provide ``suggestions'' of inputs to the automated analysis process the results were surprisingly good.

\emph{This experience suggested a shift in the current vulnerability analysis paradigm, from the existing tool-assisted human-centered paradigm to a new human-assisted tool-centered paradigm.}
Systems that follow this paradigm would be able to leverage humans (with different level of expertise) for specific well-defined tasks (e.g., tasks that require an understanding of the application's underlying logic), while taking care of orchestrating the overall vulnerability analysis process.

This shift is somewhat similar to introduction of the assembly line in manufacturing, which allowed groups of relatively unskilled workers to produce systems (such as cars) that had, until then, remained the exclusive domain of specially trained engineers.
Conceptually, an assembly line ``shaves off'' small, easy tasks that can be carried out by a large group of people, in loose collaboration, to accomplish a complex goal.

\begin{figure}[t!]
  \centering
  \includegraphics[width=1.0\columnwidth]{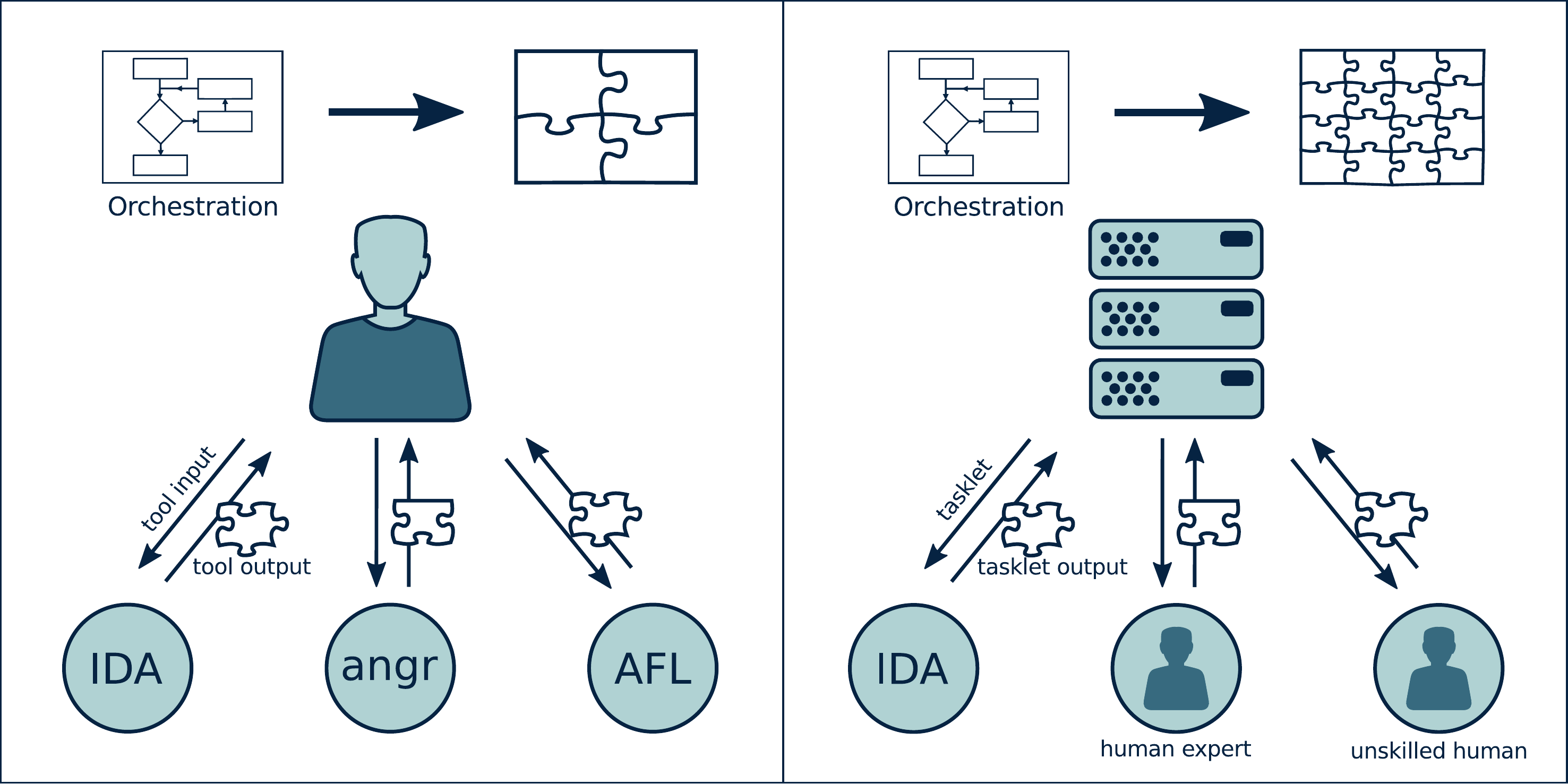}
\caption{Tool-assisted Human-centered Analysis vs. Human-assisted Tool-centered Analysis.}
\label{fig:approach}
\end{figure}

In this paper, we explore the application of this idea to vulnerability analysis.
More precisely, we develop an approach that leverages \emph{tasklets} that can be dispatched to human analysts by an autonomous program analysis system, such as those used in the Cyber Grand Challenge, to help it surmount inherent drawbacks of modern program analysis techniques (see Figure~\ref{fig:approach}).
We explore the question of how much our ``program analysis assembly line'' empowers humans, otherwise unskilled in the field, to contribute to program analysis, and we evaluate the improvement that external human assistance can bring to the effectiveness of automated vulnerability analysis\footnote{In the rest of the paper, we refer to ``automated vulnerability analysis'' as the orchestration process, even though it might include tasks that are outsourced to humans.}.
Our results are significant: by incorporating human assistance into an open-source Cyber Reasoning System, we were able to boost the amount of identified bugs in our dataset by 55\%, from 36 bugs (in 85 binaries) using fully-automated techniques to 56 bugs through the use of \emph{non-expert} human assistance.

In summary, this paper makes the following contributions:

\begin{itemize}
	\item

		We introduce the design of a \emph{human-assisted automated vulnerability analysis system}, in which the result of well-defined \emph{tasklets} that are delegated to human actors are integrated in the (otherwise) autonomous analysis process.
		These tasklets help automated analysis systems to bridge the ``semantic gap'' in the analysis of complex applications.

 	\item

 		We implemented a prototype human-assisted autonomous system on top of Mechanical Phish, a system that participated in the DARPA Cyber Grand Challenge, which was open-sourced by its authors.
		To support the community and drive the state of (semi-) automated program analysis forward, we open-source our modifications to Mechanical Phish.

 	\item

 		We experimentally evaluated the effectiveness of our tasklets in aiding the vulnerability analysis process of our system by leveraging the assistance of unskilled humans, showing that significant contribution can be made without requiring expert hackers.

\end{itemize}

In the next section, we will discuss the background of automated program analysis and pinpoint the challenges that we hope to solve with human-analyzed tasklets.

\section{Background}
\label{sec:background}

The field of vulnerability discovery has received a significant amount of research attention.
In this section, we will describe the current state of the art of both automated and manual vulnerability discovery techniques, show the challenges facing each of them, and position our approach in the context of related work.

\subsection{Fully Automated Analysis}

Individual techniques have been developed for identification of vulnerabilities~\cite{cha2012unleashing,godefroid2008automated,stephens2016driller}, automatic exploitation~\cite{huang2012crax,avgerinos2014automatic,axgen}, and automatic application protection~\cite{wang2017ramblr,wang2015reassembleable,perkins2009automatically}.
However, until recently, researchers did not focus on the integration of various techniques into cohesive end-to-end systems.
Over the last two years, DARPA hosted the Cyber Grand Challenge which required contestants to develop \emph{Cyber Reasoning Systems} (CRSes). These are fully autonomous machines capable of identifying, exploiting, and patching vulnerabilities in binary code.

A Cyber Reasoning System represents the culmination of years of research into automated binary analysis.
However, being fully autonomous, CRSes suffer from the limitations of their underlying techniques.
These limitations were reflected in the Cyber Grand Challenge results, in which only 20 out of the 87 vulnerable challenges were successfully exploited by the machine contenders~\cite{shellphish-phrack,cgc-dump}.

\subsection{Human-based Computation}
\label{sec:hbc}

While the assembly line pioneered the idea of splitting complex physical tasks (such as the assembly of a car) into small, manageable \emph{micro-tasks} as early as the 12th century~\cite{cameron1993concise}, the \emph{intellectual} equivalent was not explored until modern times.
This concept was most popularized with the Manhattan Project, in which specific computation \emph{micro-tasks} were assigned to and carried out by human ``computers''~\cite{de2013dissappearing}.
With the emergence of modern computing capability, these micro-tasks came to be chiefly carried out by machines.
As computers developed to the point where they could oversee such efforts, a formal specification of the different roles that humans and computer components can take on in computation emerged~\cite{shahaf2007towards,kosorukoff2001human,kosoruko2000social}.
This specification defines three roles:

\begin{description}
	\item[Organization Agent.]
		The organization agent is the \emph{overall intelligence}.
		It tracks the progress of work toward an overarching goal, determines what should be done, and creates micro-tasks.
		In the Manhattan Project, the organization agent was the panel of scientists leading the research effort.
	\item[Innovation Agent.]
		The innovation agent is the entity responsible for carrying out micro-tasks defined by the organization agent.
		In the Manhattan Project, the innovation agents were the human ``computers'' solving computation tasks.
	\item[Selection Agent.]
		The selection agent collates the results produced by the innovation agents and determines which are valid.
		In the Manhattan Project, this task was performed by the scientists leading the effort.
\end{description}

Systems are described using three letters, depending on whether a human or computer agent is responsible for each role.
For example, an \texttt{HCH} designation would imply a system with a human deciding which tasks to execute, a computer executing them, and the human deciding which of the results are useful.
In a security context, this might be the human specifying jobs to a symbolic execution engine, and then analyzing its output to identify exploitable bugs in a piece of software.

Over the last few years, the Internet has achieved enough saturation to support complex combinations of human and computer agents.
For example, Amazon's Mechanical Turk provides an API for automatically specifying micro-tasks for human consumption~\cite{barr2006ai}, usually used in a \texttt{CHC} context.
In fact, we use Mechanical Turk for many of our experiments in this paper.
In a similar vein to Mechanical Turk, specific-purpose platforms have been created to leverage human effort in the pursuit of a single overarching goal.
One such platform, Galaxy Zoo~\cite{willett2013galaxy}, utilizes human-completed micro-tasks for the classification of astronomical images, while another, Foldit\cite{eiben2012increased}, aids protein folding algorithms by having humans play ``folding games.''

\subsection{Human-Driven Automated Analysis}

Because it is important to understand the interactions between manual and automated processes in binary analysis systems, we provide a few examples of their intersections outside of the context of our work.

\paragraph{Fuzzing.}
Generational fuzzers, such as Peach~\cite{peachfuzz}, attempt to create inputs conforming to a specification that a program is designed to process.
Mutational fuzzers, such as AFL~\cite{afl}, mutate previously-known inputs to identify program flaws.

The most common way of creating these inputs and input specifications is \emph{manually}, through human effort.
This results in an \texttt{HCH} system -- a human creates the input specification, the computer performs the fuzzing, and a human analyzes the results.

An example of successful human-computer cooperation in binary analysis is the discovery of the Stagefright vulnerability in the Android multimedia library.
This vulnerability was found by repeating the following steps~\cite{stagefright-blackhat}:
\begin{description}
	\item[Organization - H.]
		The analyst seeds a mutational fuzzer (in this case, AFL), and starts it.
	\item[Innovation - C.]
		The fuzzer identifies vulnerabilities in the target application (in this case, the Android multimedia library).
	\item[Selection - H.]
		The human collects the vulnerabilities and \emph{fixes them} so that future iterations of the full system will identify \emph{deeper} vulnerabilities.
\end{description}

By repeating this \texttt{HCH} process, the analyst was able to identify many high-impact vulnerabilities inside the Android multimedia library, requiring multiple patches and an eventual rewrite of the entire library to fix~\cite{nougat-rewrite}.

\subsection{Human-\emph{Assisted} Automated Analysis}

The Cyber Grand Challenge required a fully autonomous system (\texttt{CCC}, by the definitions in Section~\ref{sec:hbc}).
This necessitated the development, by participating teams, of complex automation to handle the organizational, innovation, and selection roles.
However, we propose that while the organizational and selection roles must be automated to achieve high scalability, some human effort can still be used in the innovation role to mitigate drawbacks currently impacting automated program analysis techniques.
That is, our intuition is that it is possible to create a \textbf{H}uman-\textbf{a}ssisted \textbf{C}yber \textbf{R}easoning \textbf{S}ystem (\cyborg{}) that would sparingly use human assistance to improve its performance.

\cyborg{} provides a principled framework for such an integration of manual and automated analysis.
It can be modeled as a \texttt{C(C|H)C} system: it does most of its work fully autonomously, but relies on human intuition in the innovation phase, when the automated processes get ``stuck.''
In this paper, we propose that limited human assistance can be used in the scope of otherwise-automated binary analysis systems.
While this has been explored in the context of generating inputs for Android applications, it has never been investigated in the context of an other-wise autonomous Cyber Reasoning System~\cite{machiry2013dynodroid}.
In the next section, we will give an overview of our system, followed by in-depth details and an evaluation of its improvement over fully-autonomous systems from the Cyber Grand Challenge.

\section{Overview}
\label{sec:overview}

While DARPA's Cyber Grand Challenge drove the integration of cutting edge automated binary analysis techniques, it also revealed the many \emph{limitations} of these techniques.
Our work on \cyborg{} extends the concept of a Cyber Reasoning System by defining a method for human interaction that compensates for many of these limitations.
Primarily, \cyborg{} is an autonomous Cyber Reasoning System.
However, when it identifies situations that can benefit from human analysis, \cyborg{} dispatches self-contained \emph{tasklets} and assigns them to human assistants.
These human assistants can vary in skill, from abundant low-skill analysts to rare high-skill hackers.

Our HaCRS can dispatch a variety of tasklets to human assistants, depending on changing requirements.
Generally, each tasklet includes a specific program that must be analyzed and a request for specific information that the human can extract from this program.
These tasklets are created by a centralized orchestration component and disseminated to the assistant through a Human-Automation Link (HAL).
In this paper, as an initial exploration of this idea, we focus on human-assisted input generation, leaving the exploration of other tasklets to future work.

\begin{description}
	\item[The Cyber Reasoning System.]
		\cyborg{} is based on \emph{Mechanical Phish}, an open-source Cyber Reasoning System that was created by Shellphish, the hacking team of the SecLab of UC Santa Barbara, and competed in the DARPA Cyber Grand Challenge~\cite{shellphish-phrack, shellphish-cgc}.
		Shellphish designed Mechanical Phish as a set of discrete components, providing individual analysis tasks, united by a central component that handles the ``overarching intelligence''~\cite{shellphish-phrack}.
		This makes it straightforward (though, unfortunately, non-trivial) to extend Mechanical Phish with other analysis techniques, such as tasklet dispatching.

		To the interested reader, we describe the relevant design details of Mechanical Phish in Section~\ref{sec:mechaphish}.
	\item[Human-Automation Link.]
		We extend Mechanical Phish to request assistance, from non-expert humans, in principled ways.

		The prototype action that we explore in this paper is \emph{input generation}.
		In input generation, input testcases are created through both automated and human-assisted techniques to form a base set of testcases to use in vulnerability discovery.
		We describe this task, the conveyance of task-specific information in a human-friendly format, and the use of the results in our Human-assisted Cyber Reasoning System in Section~\ref{sec:assistance}.
\end{description}

Next, we will discuss relevant details of Mechanical Phish before delving into the details of our tasklets.
After this, we will evaluate human performance in the execution of these tasklets against automated alternatives derived from the state-of-the-art in program analysis.

\section{The Cyber Reasoning System}
\label{sec:mechaphish}

\begin{figure*}
	\includegraphics[width=1\textwidth]{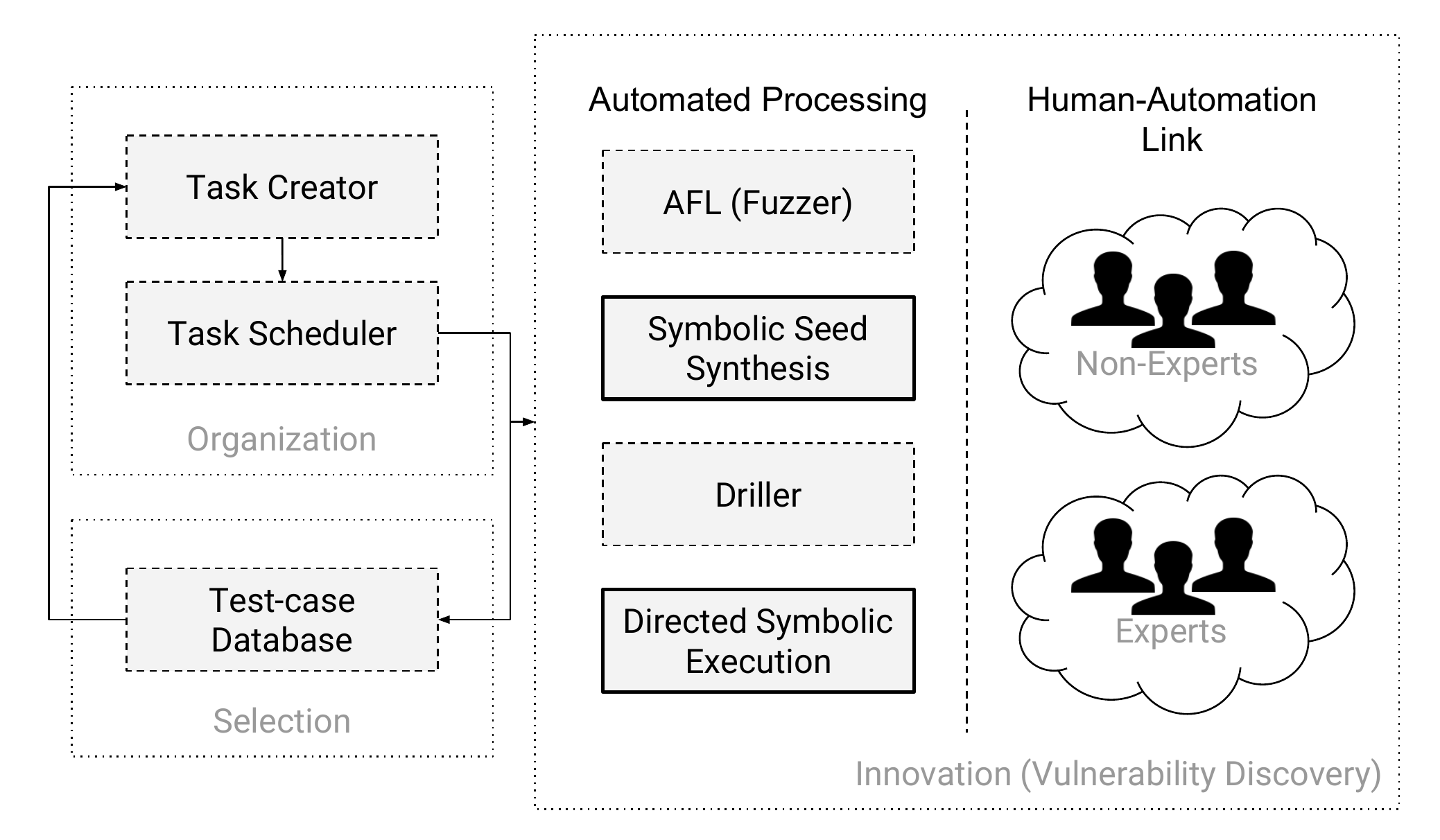}
	\label{fig:configuration}
	\caption{The HaCRS configuration. HaCRS builds upon the vulnerability discovery component of the Mechanical Phish and expands it with a Human-Automation Link to leverage non-expert human assistance in the vulnerability discovery process. Each subsystem of the vulnerability discovery component, except for the fuzzer itself, has both an automated and a human-assisted alternative. The components with a dashed border were already present in Mechanical Phish. We created the solid-bordered components for HaCRS.}
\end{figure*}

We based our implementation on the Cyber Reasoning System developed for the Cyber Grand Challenge and open-sourced by Shellphish~\cite{shellphish-cgc}.
While Mechanical~Phish is composed of modules that are spread over more than 30 different source code repositories, the core design appears to be fairly straightforward~\cite{shellphish-cgc}.

In this section, we will describe Mechanical Phish in terms of the computation framework discussed in Section~\ref{sec:hbc}.
First, we will discuss the type of software that Mechanical Phish is designed to analyze.
Then, we split the existing design into the Organization Agent, Innovation Agent, and Selection Agent, as defined in Section~\ref{sec:background}.
Afterwards, in the next section, we will detail our extensions on top of Mechanical Phish, and the specific points at which we insert human interaction.

\subsection{Program Analysis Targets}

Mechanical Phish was built for participation in the Cyber Grand Challenge.
The Cyber Grand Challenge used a custom operating system, DECREE, to ease the implementation load on participants.
To simplify analysis tasks, DECREE supports software written with a text-based interface, using seven system calls, roughly equivalent to the Linux system calls \texttt{exit}, \texttt{write}, \texttt{read}, \texttt{select}, \texttt{mmap}, \texttt{munmap}, and \texttt{getrandom}.

Aside from this simplified environment, DECREE places no restrictions on the complexity of the software itself.
As such, applications written for the Cyber Grand Challenge vary widely in complexity, from text-based video games to ``Computer-aided design`` software to web servers, and provide significant challenges to the current state-of-the-art in program analysis.
Additionally, it is important to stress that all analysis done by HaCRS takes place on \emph{binaries}, and thus functions without the semantic hints present in source code.

\subsection{Organization Agents}

The Mechanical~Phish is a \emph{state-less} Cyber Reasoning System, where, for each decision, all of the information available to Mechanical Phish, such as the binaries to be analyzed and the currently-available results of analysis components, is re-analyzed from scratch.
According to the authors, this was done in an attempt to reduce the complexity of the organizational components by freeing them from the requirement of tracking their own prior decisions~\cite{shellphish-cgc}.

Mechanical Phish includes several organizational components:

\begin{description}
	\item[Task Creator.]
		The task creator analyzes currently available results and identifies tasks that should be created, and their priorities.
		This component is actually a conglomeration of individual, task-specific creators.
		Each task-specific creator schedules its own tasks without input from other creators: the only interaction between the creators of different tasks happens when results of those tasks influence the current set of analysis results (and, in turn, are used by the subsequent tasks created by these creators).
	\item[Task Scheduler.]
		Each task is assigned a priority by its creator.
		The task scheduler analyzes task priorities and available system resources and determines which tasks to schedule.
	\item[Environment Interaction.]
		In order to inject data into Mechanical Phish, and submit the results, interaction with the environment is required.
		This component handles the retrieval of input into and exposure of output out of the system.
		While in the CGC this interaction was very straightforward, Cyber Reasoning Systems operating in other environments (for example, in a real-world cyber warfare situation) might require considerably complex agents for this task.
\end{description}

The first task that the system must carry out is the integration of environment information (for example, which binaries are available for analysis), after which the Innovation and Selection Agents can run.
	
\subsection{Selection Agents}

The selection agents are responsible for the integration of the results that are produced by the innovation agents.
However, the Mechanical~Phish does not make a distinction between the innovation agents and the integration agents in most cases.
One exception is:

\begin{description}
	\item[Vulnerability triaging.]
		When crashes are identified by the vulnerability discovery component, they are triaged to determine the feasibility of transforming them into exploits.
		This information is then used by the Task Creator to prioritize exploitation tasks based on the crash.
	\item[Exploit selection.]
		The exploits created by the Exploitation Agents are checked against different variations of the target binaries to verify that, for example, opponent systems did not patch the vulnerability.
		Successful exploits are entered into the database, to be submitted by the Environment Interaction Agent.
	\item[Patch selection.]
		Mechanical Phish implements a simple patch selection criteria, preferring patches produced by advanced (but more failure-prone) techniques than simple (but higher-overhead) ones.
\end{description}

The results of these agents are used by the organizational components to schedule further innovation tasks.

\subsection{Innovation Agents}

The tasks that are created and scheduled by the Organization Agents are carried out by the innovation agents.
Specifically, Mechanical Phish includes the following agents:

\begin{description}
	\item[Vulnerability discovery.]
		Mechanical Phish uses a combination of fuzzing and symbolic execution to analyze target binaries.
		These are implemented as separate agents that interact through cross-pollination of dynamic test cases.
		Specifically, as proposed by Driller, a coverage-based fuzzer is used in parallel with a symbolic tracing technique to produce inputs that maximize code coverage~\cite{stephens2016driller}.
	\item[Exploitation.]
		Several different exploitation agents are used by Mechanical Phish, depending on the types of vulnerabilities that are discovered.
	\item[Patching.]
		Mechanical Phish uses a complex patching agent, in several different configurations, to patch the vulnerabilities that it identifies in binary code.
\end{description}

These innovation agents process inputs and produce updates to the system state.
These updates are filtered through selection agents before the system state accepts them.

\subsection{Automated Vulnerability Discovery - Fuzzing}
\label{sec:auto-fuzzing}

The fuzzing approach in the Mechanical Phish is based on a mutational fuzzer known as American Fuzzy Lop~\cite{afl}.
This approach requires, as input, a set of test cases that exercise some functionality in the target binary.
The seed quality, in terms of how well they exercise the target program, has a scaling effect on the effectiveness of AFL: the more coverage these test cases provide, the more code AFL will be able to explore by mutating them.
Unfortunately, the creation of high-quality test case seeds is a complicated problem, and this is generally seen as a human-provided input into a system.
For example, lacking human input, Mechanical Phish simply seeds its fuzzer with an input comprised of the word ``fuzz.''

These seeds are then mutated to explore more and more of the code base and increase the chance of triggering bugs.
Eventually, however, the fuzzer will \emph{get stuck} and be unable to exercise new paths through the code of the target program.
This can happen for a number of reasons, but is most frequently caused by the inability of the fuzzer's random mutations to satisfy complex conditions, introduced by checks in the program, upon input data.

\subsection{Automated Vulnerability Discovery - Drilling}
\label{sec:auto-drilling}

Driller proposed a mitigation for the stalling of the fuzzer due to the inability to satisfy complex solutions.
It uses concolic execution to trace the paths that the fuzzer finds, identifies conditional checks that the fuzzer fails to satisfy, and synthesizes inputs to satisfy these conditions.
Driller triggers its operation when the fuzzer gets ``stuck'', and is unable to find further testcases (it detects this by checking AFL's progress evaluation heuristics).
Once this stall condition is detected, Driller symbolically traces and attempts to mutate all test cases that AFL has found into test cases that reach parts of code not previously seen.
These resulting test cases are then synchronized back into the fuzzer, so that it can explore these newly-reached areas of code.

By pairing fuzzing with concolic execution, Driller achieves better results than the naive union of the individual underlying techniques.
However, Driller's automated approach to symbolic input synthesis has some drawbacks.

Driller's synthesis works by \emph{diverting} a path and forcing it to satisfy a check that it would have otherwise avoided.
There are several limitations, inherent in Driller, that hamper its effectiveness in certain situations.
These include, but are not limited to:

\begin{description}
	\item[SMT solver.]
		Driller uses an SMT solver to solve negated \emph{path predicates} (constraints on the input values to the program that must be satisfied in order to trigger the path in question) to synthesize inputs that diverge from the original execution.
		However, depending on the complexity of the path predicates involved, the SMT solving process may not terminate.
		While this represents a significant challenge for Driller, the complexity of these predicates might not translate to the complexity of interaction with the software.
		If this is the case, a human assistant might be able to controllably divert the path taken through the program, even when the constraint solver cannot.
	\item[Inflexible path predicates.]
		Depending on implementation details in the program, earlier path predicates might prevent the deviation of later path predicates.
		Such predicates are frequently created by certain input transformation procedures.
		For example, string-to-int translation (such as the \texttt{atoi} function) takes different conditional branches, based on the values in the input string, while converting an input string to an integer.
		These conditional branches create path predicates.
		Later, the program might perform some action based on the value of this integer.
		When Driller attempts to divert this decision to take a different action, the earlier path predicates on the input string prevent this diversion.

		Humans, of course, do not share this inflexible way of reasoning about path predicates.
	\item[Semantic transitions versus control flow transitions.]
		Driller cannot understand the program semantically, and simply attempts to deviate the control flow of the program.
		A human, on the other hand, can identify much more intricate \emph{semantic} deviations (for example, winning, as opposed to losing, a game), allowing for the triggering of whole new areas of code to deal with these new semantic settings.
\end{description}

These limitations conspire to erode Driller's ability to produce deviating inputs in many cases.
In the next section, we will discuss how these limitations can be worked around with human assistance.

\section{Human Assistance}
\label{sec:assistance}

As we discuss in the previous section, automated input synthesis techniques suffer from limitations that cause them to eventually get \emph{stuck} in the exploration of a program.
Even Driller, which leverages the power of symbolic execution to divert testcases, is only a partial solution.
This is because, while Driller can make major changes to the input testcase it analyzes, it can only (by design and fundamental limitation) achieve only minor deviations.

On the other hand, a human can leverage intuition and a semantic understanding of the target program to achieve very large deviations, potentially allowing further analyses to continue to make progress.
In this paper, we explore the integration of human assistance into a Cyber Reasoning System as Innovation agents, keeping the Organizational and Selectional agents fully automated.
We focus on the vulnerability discovery stage of the analysis and explore ways to integrate human effort to improve analysis efficiency.

Human assistance takes place over an interface (the \emph{Hardware Abstraction Link}, or HAL) which will be described later this section.
To maximize the effectiveness of this effort, HaCRS carries out a number of analyses that enhance the data it is able to expose to the humans.
In this section, we describe how human assistants are selected, the interface over which HaCRS and humans communicate, and how the resulting data is used to enhance the vulnerability detection ability of HaCRS.

\subsection{Assistant Expertise}
The style of human assistance differs according to the assistant's expertise level.
For example, while HaCRS could reasonably ask an expert human to analyze a control flow graph and identify potential paths through it, a non-expert would be flabbergasted by such a request.
The information presented, and the interfaces which are used, must be adapted to the chosen assistant's level of expertise.

Since expert humans (i.e., binary analysts) are rare and expensive, the integration of assistance from non-expert humans (i.e., an average internet citizen) is of particular interest.
While they do not scale to the extent of automated processes, non-expert humans scale considerably easier than experts, due to their higher availability.
When more knowledge is required, semi-experts (i.e., undergraduates in Computer Science) can be leveraged more readily than experts.
Thus, in this paper, we focus mainly on techniques to integrate non-expert assistance, with a detour into semi-expert assistants for completion.

Over the decades that humans have been interacting with software, the skill of performing such interaction has become gradually instilled in the human population.
As such, even non-experts are well-trained to understand and drive computer software.
Thus, we can tailor HAL to non-experts by sticking to concepts that they can grasp and avoiding complex program analysis concepts, as shown in Table~\ref{tab:capabilities}.
For example, rather than ''triggering transitions'', we used the term ``triggering functionality'', which requires less technical knowledge to understand.
Additionally, we expose non-experts only to the input and output log associated with prior interactions with the programs that the HaCRS is trying to analyze, and avoid any use of program analysis terms in task descriptions.

\begin{table}
	\centering
	\small
	\begin{tabular}{l:ccc}
		\hline
		Concept & Computer & Expert & Non-Expert \\
		\hline
		\hline
		Symbolic Equations & \checkmark & & \\
		Control-Flow Graph & \checkmark & \checkmark & \\
		Execution Path & \checkmark & \checkmark & \\
		I/O (Text) & \checkmark & \checkmark & \checkmark \\
		Semantic Meaning & & \checkmark & \checkmark \\
		\hline
	\end{tabular}
	\caption{Program analysis concepts, as they are easily understood by automated techniques, expert humans, and non-expert humans. To be understandable to non-experts, the Human-Automation Link must avoid complex program analysis topics.}
	\label{tab:capabilities}
\end{table}

\subsection{Human-assisted Input Generation}

HaCRS uses human assistance to break through the ``semantic barriers'' that limit the effectiveness of automated analyses described in Sections~\ref{sec:auto-fuzzing} and~\ref{sec:auto-drilling}.
It gives its human assistants a goal: generate an input testcase that executes some amount of code in the target program that has not been reached by previously-known testcases (i.e., those previously found by automated analyses or other humans).

Human assistants interact with the target program to generate testcases, and these testcases are synchronized throughout HaCRS' components.

\begin{description}
	\item[Human-to-automation.]
		Human-produced testcases are synchronized to the automated program exploration components, which proceed to mutate them in an attempt to trigger new functionality.

	\item[Human-to-human.]
		Humans can view and modify the testcases produced by other human assistants.
		This enables a collective effort of the understanding and leveraging of program semantics toward a higher code coverage.
		HaCRS 

	\item[Automation-to-human.]
		The resulting automation-mutated testcases can then be shown to the human assistants (we term such a testcase an ``example testcase''), who can review them, understand possible further improvements and changes that can be made, and relay those changes back to the automation by producing human-modified testcases.

\end{description}

\paragraph{Testcase conversion.}
The synchronization of testcases from automated components to a human assistant poses a challenge: automated systems, driven by either random input generation or input synthesis via constraint solving, have no guarantee to produce printable characters when the target program does not require it.
Non-printable testcases look like gibberish when shown to a human, which hinders the human's ability to reason semantically about what actions the testcase is causing the target program to take.

To address this issue, we use the existing \texttt{afl-tmin} utility shipped with AFL~\cite{afl}.
This utility is a \emph{testcase minimizer}.
It takes an input testcase and uses lightweight dynamic techniques to a) remove unnecessary input characters and b) convert as many characters as possible to be printable, without changing the code coverage achieved by the input.
In practice, it achieves very good results on programs with a text interface.

\subsection{Automation-assisted Human Assistance}
\label{sec:haha}

Simply presenting previously-discovered testcases to human assistants enables an improvement over a base-case Cyber Reasoning System (we show this in Section~\ref{sec:evaluation}).
However, since the communication between HaCRS and humans takes place over a well-defined interface, HaCRS can provide extra information and capabilities to enhance the humans' abilities to complete the assistance task.

\paragraph{Interaction assistance.}
One such capability provided by HaCRS is the automated re-formatting of input data.
HaCRS traces each program testcase to detect if input data must be provided in a specific format.
It achieves this by leveraging existing techniques in protocol recovery~\cite{caballero2007polyglot,lin2008automatic,caballero2009dispatcher}.
Depending on configuration (and expertise of human assistants), this information can either be presented to the human assistants or utilized automatically to mutate human-created inputs into a format understood by the application.

In our prototype, we mainly utilize these techniques to automatically recover non-standard field delimiters used by the binaries in our dataset, but they can also be used to support information packing protocols, such as ASN1.

\paragraph{High-level guidance.}
Having enabled human interaction for binaries with complex input data specifications, HaCRS turns to the question of maximizing the ability of its humans to understand how to interact with the target program.
It does this by identifying and categorizing constant string references in the binary.

HaCRS identifies static string references by analyzing its CFG, and performs a static data flow analysis to categorizes these into strings produced as output by the program and strings compared against user input into the program.
HaCRS identifies output strings by detecting when they are passed into common output functions (such as \texttt{puts} and \texttt{printf}).
Input strings are considered to be anything that is passed to a string comparison function.
In the case of statically-linked binaries, HaCRS can leverage the function identification functionality built into the Mechanical Phish, which detects common functions in programs using pre-defined dynamic test-cases~\cite{shellphish-phrack}).

HaCRS provides a list of potential output strings in the target program to help its human assistants, relaying which of these strings have not yet been triggered (i.e., caused to be output by the program) by other testcases.
These can provide useful semantic information regarding the untapped functionality of the target program.

While HaCRS focuses on text-based software, it is important to keep in mind that analogous information can be recovered for software with a graphical user interface.
For example, a similar analysis can identify GUI widgets, render them, and display them as potential directions of exploration for human assistants.

\paragraph{Symbolic tokens.}

First, HaCRS creates suggestions for human assistants for ways that testcases might be modified to divert program flow.
This is done through a process of \emph{symbolic tokenization}.
HaCRS symbolically traces the target program in the context of each testcase to recover constraints placed on the input by the target program.
It analyzes these constraints to identify contiguous bytes on which the constraints are \emph{similar} (in terms of the number of constraint expressions and the types of arithmetic and boolean operations the constraint expressions are composed of).
These contiguous bytes represent tokens processed and reasoned about by the binary.

HaCRS then identifies alternate values for each symbolic token.
It rewinds its symbolic trace to the address at which the first constraint of the token was introduced, and performs a symbolic from that point to retrieve other potential values.
The symbolic exploration runs until a timeout (we found 30 seconds to be a reasonable timeout in our experiments).
At the end of the timeout, the constraints of the various resulting paths are solved to alternate values for the token.
These values are further refined by matching them against the input strings retrieved previously, and HaCRS produces two different sets of suggestions to its assistants: ``educated guesses'', which are the input strings that are prefix-matched by the recovered alternatives and ``brute-force guesses'', which are the raw alternatives themselves.

Note that, while the concept of generating alternatives for input is shared with Driller, the goal is different.
Driller generates alternative testcases to drive execution down different paths.
However, the alternatives generated by this method are meant to be learned by humans, understood, and reasoned about to produce new inputs through human intuition and previously-learned experience.

\paragraph{Input type annotation.}
Programs process different inputs differently, and HaCRS exposes this to its assistants by highlighting input bytes that are constrained by similar constraints (as with the symbolic token analysis, we use constraint count and operation types to compute constraint similarity).
Input bytes highlighted with similar colors in the input testcases will be bytes that have been treated similarly to each other by the program, and may represent similar type of data.
Most importantly, this differentiates string input (such as a command) against numeric input (which is passed to functions such as \texttt{atoi}, which impose specific constraints on the data).

\subsection{Human-Automation Link}
\label{sec:hai}
\begin{figure}
        \centering
        \includegraphics[width=1\columnwidth]{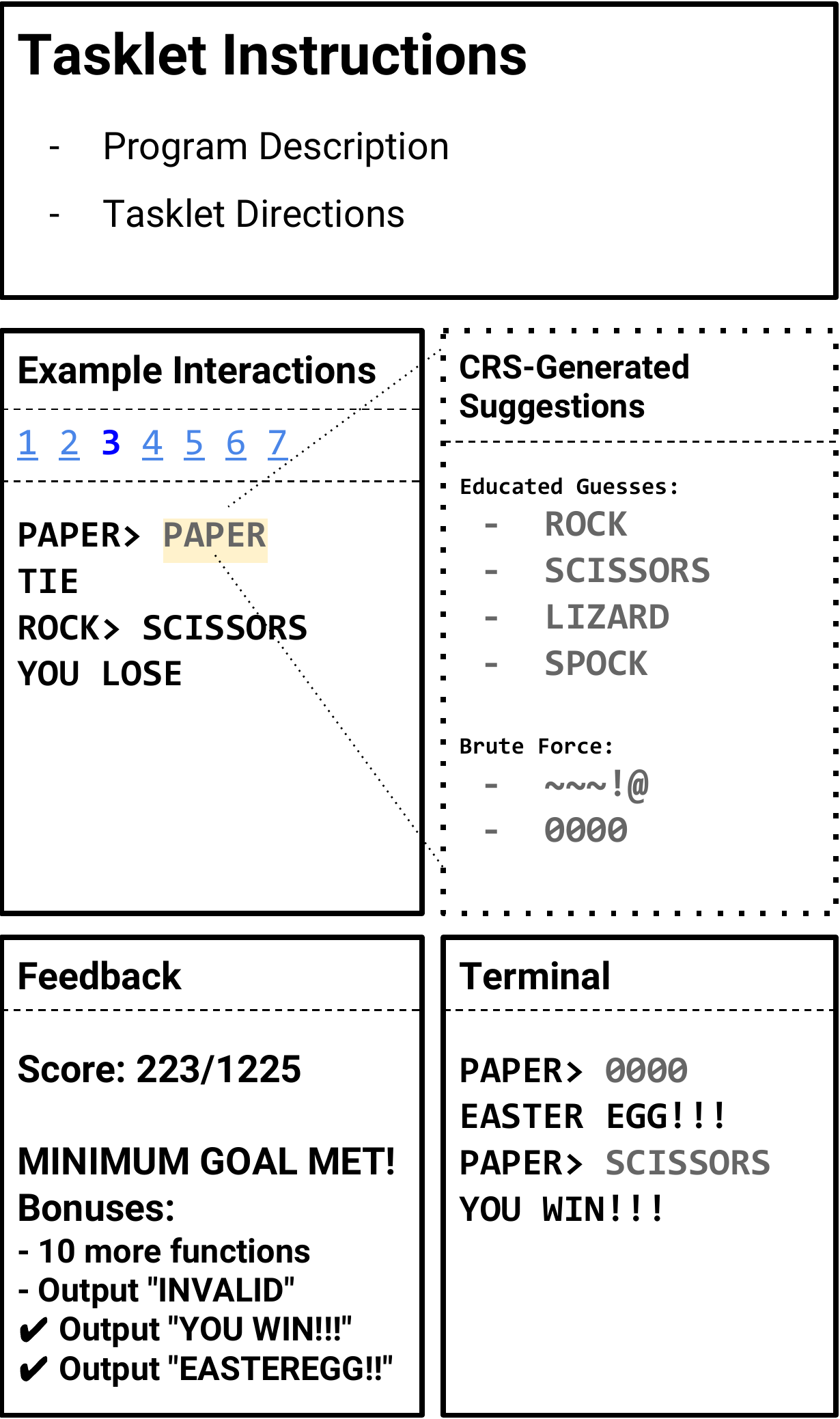}
        \caption{A diagram of the HaCRS user interface, showing different components for seeding tasklets
 (all solid-bordered components), drilling tasklets (dotted-bordered components), and seeking tasklets (d
ashed-bordered components).}
        \label{fig:interface}
\end{figure}

The interface between the HaCRS and its human assistants must be designed in such a way as to be understandable by both parties.
To do this, we created a Human-Automation Link (HAL) that exposes, to the humans, only the concepts of program analysis that non-experts might be familiar with.
For the curious reader, we reproduce a mock-up of the HAL interface in Figure~\ref{fig:interface}.

The HAL interface in Figure~\ref{fig:interface} consists of the following elements:

\begin{description}
	\item[Program description.]
		When a description of the target program is available, it can aid assistants in interacting with it.
		In the case of Cyber Grand Challenge binaries, this description includes a very brief (usually four to five sentences) summary of the program's purpose, as written by the program authors.
		In a real-world setting, human assistants can be provided with the technical manual for the piece of software being tested.
	\item[Tasklet instructions.]
		The HaCRS provides human-readable instructions, which are presented to the assistant alongside each tasklet.
	\item[Example interactions.]
		The HaCRS provides logs of previous interactions with the software, in the form of input and output data. %
		For the text-based software of DECREE OS, to help assistants understand what data was originated from them (program input) and what came from the program (program output), the input and output are displayed in different colors.
		A version of HaCRS for software with a graphical user interface could instead have a video record of the interaction, but this is not supported by our prototype.
	\item[CRS-Generated Suggestions.]
		To help assistants understand how to deviate from a test case, they can invoke the deviation annotation interface.
		This interface displays data recovered through the automated analyses described in Section~\ref{sec:haha} to present the assistant with a better idea of how to make a program behave differently than in the example testcase.
	\item[Interaction terminal.]
		To facilitate the interaction between human assistants and the target program, a terminal is presented to interact with the software.
		Again, to help assistants understand differentiate user input from program output, the input and output are displayed in different colors.
	\item[Tasklet goal and feedback.]
		Any human-facing task must have an understandable end goal to avoid confusion on the part of the assistants.
		HaCRS requires its human assistants to trigger previously-unseen functionality in the target programs.
		To this end, it provides feedback to the assistant regarding the amount of previously-unseen control flow transitions that the assistant was able to trigger.

		Along with this, it provides a display of \emph{untriggered output strings}, as described in Section~\ref{sec:haha}.
		With their human ability to reason about semantic information, assistants can leverage the bounty strings to better target untriggered functionality in the program.
\end{description}

Each tasklet also has a timeout and an abort button: if the assistant is unable to complete the tasklet before a timeout, or presses the abort button, the tasklet is terminated.
This acts as a guard against the situation when the tasklet is not actually completable (for example, if the remaining untriggered functionality is dead code).

In the next section, we will explore the implication of human assistance by evaluating the performance of HaCRS against the performance of the unaided Mechanical Phish.

\section{Evaluation}
\label{sec:evaluation}

In this section, we evaluate the impact produced by our integration of non-expert human effort into the Cyber Reasoning System paradigm.
We measure the result of this as a whole, in terms of the overall number of vulnerabilities identified in our dataset, but also explore certain low-level details of the system.

\subsection{Dataset}
As previously mentioned, Mechanical Phish was designed to operate on binaries for DECREE, the operating system designed for the DARPA Cyber Grand Challenge.
A total of 250 binaries were produced by DARPA for the Cyber Grand Challenge\footnote{DARPA recently funded the creation of a human-readable repository with information on these applications, hosted at \texttt{http://www.lungetech.com/cgc-corpus}.}.
These binaries vary in complexity, but are designed to mimic a wide range of vulnerabilities and behaviors found in real-world software.
Each Cyber Grand Challenge binary is guaranteed to have at least one vulnerability, and proof-of-concept exploits, along with high-quality testcases, are provided for each.
This makes it possible to measure, with some degree of certainty (after all, previously-unknown vulnerabilities might also be present), the effectiveness of vulnerability detection techniques.
As such, they have already been used in the evaluation of various other scientific work~\cite{shoshitaishvili2016sok,stephens2016driller,wang2017ramblr}.

Our dataset is the subset of DECREE programs that present a human-usable text protocol or for which the interaction assistance provided by HaCRS (as discussed in Section~\ref{sec:haha}) was able to facilitate a human-usable text protocol.
We selected these by automatically detecting the presence of non-printable characters in the author-provided testcases (we did not otherwise use these testcases in the experiments).
We filtered binaries in this way because, to our human assistants, such protocols are understandable, and, therefore, they allow for manual interaction.
Among the CGC binaries, a total of \binarycount{} binaries meet this criteria.

While this requirement to filter the dataset to binaries designed for human interaction is limiting, certain approaches do exist to alleviate it.
For example, IARPA funded a multi-year effort, dubbed STONESOUP~\cite{stonesoup} that developed a number of approaches to \emph{gamify} software.
Such approaches can be used to expand the amount of binaries with which humans can assist, but they generally fail to recreate the valuable semantic hints in software designed for humans.
We leave the integration of such program mutation into our interaction assistance component as future work.

Even though a protocol might be text only, it might still be hard for humans to understand.
As an example of this, consider PDF, which is a text-only file format that is designed to be parsed exclusively by computer programs.
To better understand the implications of human assistance on the binaries in our dataset, we manually categorized them according to the following qualities:

\begin{description}
	\item[Technical expertise.]
		We determined whether a program requires technical expertise to be used.
		For example, some of the programs in the dataset are language interpreters or databases, requiring users to be familiar with such Computer Science concepts as programming languages.
		These programs would be rated as requiring high technical expertise.
	\item[Semantic complexity.]
		We attempted to identify whether actions taken by the program yield themselves to high-level reasoning about the program's \emph{intent}.
		For example, a move taken in a chess match would have high semantic complexity, whereas an iteration of a compression algorithm would not.
		Thus, a chess engine would be ranked as having high semantic complexity, whereas a compression utility would not.
\end{description}

CGC binaries are fairly small, and the small size of these binaries makes them well-suited for such classification.
Specifically, because the binaries tend to be ``single-purpose'' (i.e., a recipe storage application, as opposed to a web browser), most binaries do not have different modules with different semantic complexity or technical expertise requirements.

The binaries, by their various classifications, are presented in Table~\ref{tab:stats}.
We expect human assistants to do best on binaries with a high semantic complexity, and unskilled humans to do best with binaries requiring a low technical expertise.

\begin{table*}
	\begin{tabular}{ll|llll}
		\hline
		\pbox{18mm}{\strut Semantic Complexity \strut} & \pbox{15mm}{\strut Technical Expertise \strut} & Binaries \\
		\hline
		\hline
		High & Low &
		   \texttt{CADET\_00001 CADET\_00003 CROMU\_00001 CROMU\_00003 CROMU\_00005 CROMU\_00017} \\
		&& \texttt{CROMU\_00029 CROMU\_00031 CROMU\_00037 CROMU\_00040 CROMU\_00041 CROMU\_00044} \\
		&& \texttt{CROMU\_00046 CROMU\_00054 CROMU\_00065 CROMU\_00076 CROMU\_00087 EAGLE\_00005} \\
		&& \texttt{KPRCA\_00011 KPRCA\_00017 KPRCA\_00018 KPRCA\_00022 KPRCA\_00023 KPRCA\_00026} \\
		&& \texttt{KPRCA\_00030 KPRCA\_00042 KPRCA\_00043 KPRCA\_00049 KPRCA\_00051 KPRCA\_00052} \\
		&& \texttt{KPRCA\_00053 KPRCA\_00055 KPRCA\_00071 KPRCA\_00079 NRFIN\_00004} \\
		&& \texttt{NRFIN\_00005\textsuperscript{+} NRFIN\_00065 TNETS\_00002 YAN01\_00001} \\
		\hdashline
		High & High & 
   		   \texttt{CROMU\_00002 CROMU\_00008 CROMU\_00009 CROMU\_00010 CROMU\_00011 CROMU\_00014} \\
		&& \texttt{CROMU\_00015 CROMU\_00021 CROMU\_00022 CROMU\_00023 CROMU\_00035 CROMU\_00042} \\
		&& \texttt{CROMU\_00048 CROMU\_00051 CROMU\_00071 CROMU\_00083 CROMU\_00096 CROMU\_00098} \\
		&& \texttt{KPRCA\_00007 KPRCA\_00013 KPRCA\_00021 KPRCA\_00028 KPRCA\_00031 KPRCA\_00036} \\
		&& \texttt{KPRCA\_00041 KPRCA\_00045 KPRCA\_00054 KPRCA\_00068 LUNGE\_00002} \\
		&& \texttt{NRFIN\_00001\textsuperscript{+} NRFIN\_00009\textsuperscript{+} NRFIN\_00054 NRFIN\_00055 YAN01\_00002} \\
		&& \texttt{YAN01\_00007 YAN01\_00011} \\
		\hdashline
		Low & Low & \texttt{NRFIN\_00008\textsuperscript{*} NRFIN\_00064 NRFIN\_00069\textsuperscript{+} YAN01\_00015} \\
		\hdashline
		Low & High & \texttt{CROMU\_00025\textsuperscript{*} CROMU\_00030\textsuperscript{*} CROMU\_00034\textsuperscript{*} KPRCA\_00010\textsuperscript{*} KPRCA\_00064\textsuperscript{*}} \\
		&& \texttt{NRFIN\_00071} \\
		\hline
	\end{tabular}
	\caption{The binaries in our dataset, grouped by semantic complexity of their operation and the required technical (Computer Science) expertise. These binaries were filtered for receiving mostly printable input, but some of them (marked with \texttt{*}) decoded that into raw binary input, making them suboptimal for human interaction. Others (marked with \texttt{+}) received their inputs in protocols which were automatically translated by HaCRS' interaction assistance layer to be easily human-interactable. We expect humans to do best on binaries with a high semantic complexity, and unskilled humans to do best with binaries requiring a low technical expertise.\label{tab:stats}}
\end{table*}

\subsection{Human Assistants}
HaCRS was designed to support different levels of assistant expertise, from non-experts to experts.
We evaluated the impact of both non-expert and semi-expert assistants.

\paragraph{Non-experts.}
For the non-experts, We used Amazon's Mechanical Turk service to dispatch tasklets to humans with no required Computer Science knowledge~\cite{barr2006ai}.
This provided HaCRS with an API to interact with human intelligence in a scalable way, allowing it to submit tasklets, as Mechanical Turk Human Intelligence Tasks (HITs), without concerning itself with human availability.

Because we had finite funds for our experiments, we implemented a \emph{human interaction cache}.
When the HaCRS would create tasklets for non-expert human assistance, we would first check the interaction cache to determine if this human assistance task had already been requested in by a prior experiment.
If it had, and if at least one of the cached human testcases ``solved'' the tasklet (in the sense of triggering new code), the HaCRS would reuse it instead of paying for a HIT.
We used the human interaction cache whenever we were running experiments on identical configurations of the Hardware-Automation Link.
This allowed us to re-run some of the experiments throughout the design and development of the system and remain within our budget.

In the end, between the different experiments to fully understand our system, we spent about \$2000 on Mechanical Turk HITs, resulting in 21268 unique testcases across our experiment.
While this is a large amount for a research lab, it would be trivial spending for a nation state or large corporation looking to scale out their analyses.

\paragraph{Semi-experts.}
We recruited five undergraduate Computer Science students, familiar with programming topics but not with program analysis, to act as our semi-expert human assistants.
These undergraduates interacted with a random sampling of 23 binaries from our dataset, generating a total of 115 testcases.

\paragraph{Ethics.}
As our experiments involve human assistants, we were careful to ensure that ethical procedures were followed.
We worked with the Institutional Review Board of our institution to evaluate our testing protocol.
The IRB approved our experiments, and we were careful to follow ethical guidelines throughout our work.

\subsection{Human-Automation Link}
As we proposed a number of optimizations to the Human-Automation Link in Section~\ref{sec:haha}, it is important to understand whether this actually enhances the effectiveness of human assistances.
To determine this, we performed two separate experiments in having non-experts interact with programs in the HAL, with our optimizations in Section~\ref{sec:haha} disabled in the first and enabled in the second.

For each binary, we dispatched tasklets to the human assistants until they were unable to make further progress in code coverage, given an hour-long timeout.
We collated the results by the semantic complexity of the binaries involved, and computed the median number of testcases at which progress stopped being made.

Our improvements to the HAL allowed our assistants to contribute a significantly higher amount of testcases than they were previously able to.
For semantically complex binaries, the number of testcases was roughly double, but for binaries that were not semantically complex, the improvement was considerably higher, approach a three-fold increase in the number of successful testcase generations.
On further investigation, this makes sense -- analyzing the testcases generated by the human assistants, we were able to see them quickly guess how to interact with semantically-complex programs, but struggle with less complex ones.
However, with the improved HAL interface, they were given extra information that they could leverage to provide high-quality testcases.

\subsection{Comparative Evaluation}

\begin{table*}
	\begin{tabular}{l:cc|cccccc}
		\hline
		\pbox{20mm}{Configuration} &
		\pbox{15mm}{Semantic Complexity} &
		\pbox{12mm}{Expertise Required} &
		\pbox{14mm}{\strut Median Code Coverage \strut} &
		\pbox{10mm}{Median \#AT} &
		\pbox{10mm}{Median \#HT} &
		\pbox{10mm}{Binaries Crashed} &
		\pbox{12mm}{Median Time-to-Crash} \\
		\hline
		\hline
		Non-expert Humans               & High & Low  &    46.68\% &     0 &   137 &     0 &  N/A \\
		                                & High & High &    48.83\% &     0 &   150 &     0 &  N/A \\
		                                & Low  & Low  &    48.69\% &     0 &   168 &     0 &  N/A \\
		                                & Low  & High &    16.81\% &     0 &   297 &     0 &  N/A \\
		                                \cdashline{2-8}
		                                & &  Total &    47.19\% &     0 &   151 &     0 &  N/A \\
		\hdashline
		All Humans                      & High & Low  &    46.83\% &     0 &   137 &     0 &  N/A \\
		                                & High & High &    48.83\% &     0 &   150 &     1 &  2815 \\
		                                & Low  & Low  &    48.69\% &     0 &   168 &     0 &  N/A \\
		                                & Low  & High &    17.39\% &     0 &   298 &     0 &  N/A \\
		                                \cdashline{2-8}
		                                & & Total &    47.19\% &     0 &   151 &     1 &  2815 \\
		\hdashline
		Unassisted Fuzzing              & High & Low  &    41.82\% &   410 &      0 &  12 &  807 \\
		                                & High & High &    43.32\% &   526 &      0 &  14 &  1278 \\
		                                & Low  & Low  &    56.17\% &   187 &      0 &  1  &  143 \\
		                                & Low  & High &    17.46\% &   211 &      0 &  1  &  7 \\
		                                \cdashline{2-8}
		                                & & Total &    42.87\% &   361 &      0 &  28 &  897 \\
		\hdashline
		Symbolic-assisted Fuzzing       & High & Low  &    42.90\% &   663 &      0 &  14 &  1302 \\
		                                & High & High &    48.85\% &   764 &      0 &  17 &  1426 \\
		                                & Low  & Low  &    56.07\% &   156 &      0 &  2  &   62 \\
		                                & Low  & High &    41.88\% &  1500 &      0 &  3  &   390 \\
		                                \cdashline{2-8}
		                                & & Total  &    44.91\% &   649 &      0 &  36 &  1298 \\
		\hdashline
		Human-assisted Fuzzing          & High & Low  &    49.70\% &   326 &    136 &  21 &  1378 \\
		                                & High & High &    60.45\% &   472 &    126 &  24 &  1442 \\
		                                & Low  & Low  &    64.03\% &   125 &     35 &  2  &  48 \\
		                                & Low  & High &    17.46\% &   207 &      9 &  1  &  10 \\
		                                \cdashline{2-8}
		                                & & Total  &    52.38\% &   308 &     84 &  48 &  1334 \\
		\hdashline
		Human-assisted Symbolic-        & High & Low  &    48.98\% &   369 &     69 &  23 &  1140 \\
		assisted Fuzzing                & High & High &    59.68\% &   485 &     11 &  28 &  1855 \\
		                                & Low  & Low  &    64.03\% &   121 &     46 &  2  &  47 \\
		                                & Low  & High &    48.52\% &   641 &      5 &  3  &  584 \\
		                                \cdashline{2-8}
		                                & & Total  &    53.45\% &   403 &     34 &  56 &  1301 \\
		\hline
	\end{tabular}
	\caption{The crashes found and code coverage achieved by different configurations of the automated and human components of HaCRS. The full HaCRS configuration includes human non-expert, human semi-expert, and automated innovation agents. \#AT, and \#HT are the numbers of automation-originated testcases and human-originated testcases, respectively, that were deemed ``unique'' by the Mechanical Phish's testcase evaluation criteria.}
	\label{tab:numbers}
\end{table*}

HaCRS improves the vulnerability detection process by injecting human intuition into the Cyber Reasoning System.
To understand how effective this is, we analyze the impact that non-expert and semi-expert assistance has on CRS effectiveness.
To explore these questions, we ran several different experiment configurations:

\begin{description}
	\item[Non-expert humans.]
		As a baseline to understand the ability of humans to generate inputs for binary code, we disabled the automated components of the Mechanical Phish and relied solely on human assistants for testcase creation.
	\item[Semi-expert and non-expert humans.]
		With the amount of semi-experts at our disposal, it did not make sense to have them work alone.
		As such, we ran an integrated semi- and non-expert experiment.
		To understand the impact of expertise, we added the semi-experts to our assistant pool and reran the human-only experiment.
		Testcases produced by non-experts are presented to semi-experts as examples, and testcases created by the semi-experts are synchronized into the system and eventually presented to the non-experts.
	\item[Unassisted fuzzing (AFL).]
		This configuration, with both symbolic and human assistance disabled, achieves a baseline for comparing the other experiments to understand the relative gains in code coverage and crashes.
	\item[Symbolic-assisted fuzzing (Driller).]
		This is the reference configuration of the Mechanical Phish: a fuzzer aided by a dynamic symbolic execution engine, as proposed by Driller.
		We consider this as the prior state-of-the-art configuration.
	\item[Human-assisted fuzzing.]
		In this configuration, Driller is replaced with our Human-Automation Link.
		Rather than symbolically tracing fuzzer-generated testcases, we present them to our human assistants and synchronize their testcases back into the fuzzer.
		This configuration, together with the Driller and AFL configurations, allow us to understand the relative effectiveness of Drilling versus Human Assistance.
	\item[Human-assisted Symbolic-assisted fuzzing.]
		This is the ``complete'' configuration of HaCRS, all components, representing the new state-of-the-art in Cyber Reasoning System.
\end{description}

The results of the experiment are presented in Table~\ref{tab:numbers}.

\paragraph{End-to-end system.}
The most obvious result is the improvement in the number of vulnerabilities that were identified with the full HaCRS configuration.
By iteratively combining human assistance and symbolic assistance to its internal fuzzer, the HaCRS was able to identify an additional \emph{twenty} bugs in different binaries over symbolically-assisted fuzzing (a whopping 55\% improvement) and \emph{twice} as much as the base-case fuzzer alone.
This result is significant: non-expert humans, overwhelmingly likely to have no security or program analysis training, are able to make real contributions toward the analysis of binary software.

\paragraph{Comparison to Driller.}
In HaCRS, human assistants take on a very similar role to Driller: they provide extra inputs that the fuzzer can leverage to avoid stalling in its exploration of the target program.
Rather than making small control-flow diversions, human assistants make \emph{semantic} divergences based on their understanding of the operation of the target program.
This is reflected in the results -- for semantically-complex programs, the human assistants significantly beat out Driller, achieving an improvement of up to 11.6\% improvement in coverage.
However, for binaries that did not have semantic complexity but required computing expertise, the human assistants suffered, being unable to understand the concepts presented by the program and intuit how to interact with it.
This is where the combination of human and automated analysis shines -- Driller picks up the slack in these binaries, and the combination of human and symbolic assistance achieves higher code coverage than either alone.

\paragraph{Impact of expertise.}
Interestingly, the inclusion of semi-experts in our analysis did not seriously impact the achieved code coverage.
This is an example of the different scale achievable for experts and semi-experts.
While we were able to get just over 300 Mechanical Turk workers to assist HaCRS, we were only able to recruit five undergraduate students, and they could not make a strong impact on the results (in fact, because the results are presented in aggregate, there was almost no impact on the median measurements).
However, they did have localized success: due to their ability to intelligently interact with more complex binaries, the experts were able to identify a bug in one of the applications without any human assistance.
Specifically, they triggered a bug in \emph{CROMU\_00021}, which implements a simple variable storage and arithmetic engine, but contains an exploitable bug when a variable with a blank name is created.

\subsection{Case Studies}

In the course of our experiments, our human assistants achieved some results that are interesting to explore more in-depth.
This was despite the fact that the human assistants were completely unskilled in program analysis, and were recruited with absolutely no training.
Here, we delve deeper into these bugs, and discuss why human effort helped with these specific binaries.

\paragraph{Coverage case study: CROMU\_00008.}
This binary implements a database with a SQL-inspired interaction interface.
Proper use of this binary required understanding the concepts of storing and retrieving data records.
Interestingly, our human assistants quickly developed an understanding for how to do this, taking the suggested keywords from the CRS suggestions and combining them into expressions the program understood.
They achieved a code coverage of 55.5\%, compared with 12.1\% for the automated analyses.
Manual investigation into the delta between automation and human assistance revealed that, as expected, the humans produced inputs that were meaningful for the program, while the symbolic seed synthesis attempted to optimize for code coverage, triggering many meaningless states (such as incorrect commands) without ever getting to the actual operation of the program.

\paragraph{Coverage case study: KPRCA\_00052.}
This binary is surprisingly complicated: the assistant is presented with a pizza order menu system.
To properly navigate this system, the assistant must understand how a pizza is made: the crust is chosen first, then the cheese, then the toppings.
This makes it very hard for the automated system to explore this binary and, in fact, our automation achieved a 19\% code coverage over the course of the experiment, as opposed to 52\% achieved by human assistants.

\paragraph{Vulnerability-detection case study: KPRCA\_00043.}
This binary includes a lyrical storage engine that disallows certain words from being provided as lyrics.
Furthermore, these words are checked by a \emph{bloom filter}.
Because this filter is implemented as a hash map, the resulting symbolic memory references make it difficult for symbolic execution to produce these words.
The vulnerability consists of an overflow in the lyrics buffer if enough words are entered that trigger the bloom filter but then pass the secondary check against the blacklist.
For example, enough misspellings of the blacklisted words can overflow the filter.

Interestingly, the binary includes a semantic hint -- a depressing message is printed when the program starts.
We observed that our assistants quickly picked up on this hint and produced inputs containing blacklisted words, whereas the symbolic seed synthesis produced gibberish and failed to trigger the blacklist.
While both approaches actually had a similar level of code coverage, we hypothesized that the human-assisted inputs would be a better seed set for a fuzzer to find the vulnerability.
We verified this by running AFL for an hour with the human-provided seeds, and for an hour with the automation-created seeds.
As expected, the former triggered the vulnerability, while the latter did not.

\paragraph{Vulnerability-detection case study: NRFIN\_00055.}
This binary implements a complex mathematical calculator.
In this case, the assistants were given no guidance -- simply an input prompt.
Over time, they figured out that mathematical expressions led to the output of mathematical solutions, and their apparent knowledge of mathematics came into play.
Assistants submitted a number of equations, exploring different operations in semantically-valid inputs (i.e., proper use of parenthesis and so forth).
By comparison, the symbolic seed synthesis became overwhelmed by the number of paths in the parsing engine, resulting in an inability to produce reasonable inputs.

Again, we suspected that the seeds provided by humans would be more useful to a fuzzer, and tested this by an hour-long execution of AFL.
As with the previous case, AFL was able to mutate the human-produced testcases into a crashing input by introducing a modifier symbol missed by the assistants into one of their testcases, resulting in a type confusion.
The automatically-generated testcases were less useful, and no crash was found from them within an hour.
\section{Conclusion}

The use of principled human-assistance in Cyber Reasoning Systems constitutes a paradigm shift in our view of how binary analysis is done.
Instead of the dichotomy between human-led, semi-automated systems (HCH, as discussed in Section~\ref{sec:background}) and fully automated systems (CCC), we propose a C(H|C)C system, where computers, which scale beyond human ability, make organizational calls and humans, whose intuition has not yet been replicated, assist when able.
This system can utilize the insight of \emph{non-expert humans}, who are more abundant than expert humans and thus scale better.
In the absense of these humans, these systems are able to operate fully autonomously, just at a lower effectiveness.

In this report, we have taken a first look at how non-experts impact the automated vulnerability discovery pipeline. %
The results are significant: humans, with no security training, were able to seriously improve the bug detection rate of a state-of-the-art vulnerability analysis engine.
Further exploration is warranted.
For example, humans can confirm or repudiate results of static analysis, combine behavior observed in different testcases into one, and help verify automatically-generated patches.
All of this is challenging or simply infeasible with modern techniques, but the use of human assistance can greatly augment Cyber Reasoning Systems with these capabilities regardless.

\bibliographystyle{ACM-Reference-Format}
\bibliography{biblio}

\end{document}